\documentclass[aps,twocolumn]{revtex4}
\usepackage{amsfonts,amssymb,amsmath}            
\usepackage[]{graphics,graphicx,epsfig}            
\usepackage{amsthm}                              
\def\identity{\leavevmode\hbox{\small1\kern-3.8pt\normalsize1}}

\newtheorem{definition}{Definition}
\newtheorem{lemma}{Lemma}
\newcommand{\ket}[1]{\left | #1 \right\rangle}
\newcommand{\bra}[1]{\left \langle #1 \right |}

\newcommand{\smallfrac}[2][1]{\mbox{$\textstyle \frac{#1}{#2}$}}

\newcommand{\proj}[1]{\ket{#1}\bra{#1}}
\renewcommand{\epsilon}{\varepsilon}
\begin{document}

\title{A QMA-Complete Translationally Invariant Hamiltonian Problem\\ and the Complexity of Finding Ground State Energies in Physical Systems}
\date{\today}

\author{Alastair \surname{Kay}}
\affiliation{Centre for Quantum Computation,
             DAMTP,
             Centre for Mathematical Sciences,
             University of Cambridge,
             Wilberforce Road,
             Cambridge CB3 0WA, UK}
             
\begin{abstract}
Here we present a problem related to the local Hamiltonian problem (identifying whether the ground state energy falls within one of two ranges) which is restricted to being translationally invariant. We prove that for problems with a fixed local dimension and $O(\log(N))$-body local terms, or local dimension $N$ and 2-body terms, there are instances of the problem which are QMA-complete. We discuss the implications for the computational complexity of finding ground states of these systems, and hence for any classical approximation techniques that one could apply including DMRG, Matrix Product States and MERA. One important example is a 1D lattice of bosons with nearest-neighbor hopping at constant filling fraction i.e.~a generalization of the Bose-Hubbard model.
\end{abstract}

\maketitle

{\em Introduction:} The complexity class QMA is the quantum analogue of NP i.e.~it is the class of problems for which there exists a polynomially sized quantum computation $V$ that can verify the solutions to the problem, outputting `yes' if the input is indeed a solution, and `no' otherwise. Recent interest has focused around establishing complete problems for QMA, as these encapsulate the character of the class. This started with Kitaev's proof that $O(\log N)$-local and 5-local Hamiltonian problems are QMA-complete \cite{KSV02a}, where a Hamiltonian problem is defined to be establishing whether a particular Hamiltonian has its smallest eigenvalue less than one threshold, $a$, or greater than another, $b$, where $b>a$. 
Since this original proof, the structure of Hamiltonians which still solve QMA-complete problems has been reduced to 2-local Hamiltonians \cite{Kempe}, and their spatial organisation in 2D \cite{oliveira-2005} and 1D lattices \cite{gottesman} has also been demonstrated. In the one-dimensional case, this was achieved by increasing the local dimension of spins from 2 to 12. Reductions of other, seemingly unrelated, problems have also been shown, hence proving their QMA-completeness, such as the $N$-representability problem \cite{liu-2007-98}.

In this paper, we impose the translational invariance over a one-dimensional (periodic) chain on the Hamiltonian, which can elucidate the trade-offs in resources required for quantum computation. One extreme that we already know is given by the aforementioned local Hamiltonian constructions \cite{kay-2007}, where we have no time resolution in our control of the system, only spatial resolution. Another extreme is where we exchange the spatial control for temporal control \cite{vollbrecht-2005}, which is the culmination of the study of global control schemes (see \cite{kay-2007a} and references therein). The problem that we develop here, attempting to find a translationally invariant Hamiltonian for computation, has neither spatial nor temporal control. Consequently, there is a cost elsewhere in the scheme - either in the local dimension of the spins used ($\sim\text{poly}(N)$ where there are $N$ qubits), or the range of the Hamiltonian terms ($\sim\log(N)$). 

The ultimate aim of this study, however, is to examine the computational complexity of finding the ground state energy of a translationally invariant local Hamiltonian. It is typically argued that since there are exponentially many parameters required to define a state, determining the ground state of an arbitrary Hamiltonian is a hard problem. However, by restricting the symmetry of a system, the effective number of parameters is reduced and thus one might hope to efficiently approximate the ground state of, say, translationally invariant local Hamiltonians. This is a hugely important task in condensed matter physics since these are the Hamiltonians that typically occur in nature, and is precisely the problem that is tackled by a range of approximation techniques such as DMRG/Matrix Product States (MPS) \cite{Whi92a,Whi93a,DMRG_period}, Quantum Monte Carlo simulations \cite{RevModPhys.73.33} and MERA \cite{vidal-2006a}. There are good reasons to expect that efficient representations of these states exist \cite{verstraete-2006-73,hastings-2006-73}, and so it becomes a question of the computational difficulty of finding a good representation. One strikingly open question is ``how accurately can the ground state of a one-dimensional translationally invariant Hamiltonian and the corresponding energy be approximated given polynomial time?", and it is hoped that one could discover techniques where it is possible to present certificates that guarantee a minimum level of accuracy. Some steps have been taken in understanding this problem, providing NP-hard instances of MPS \cite{eisert-2006c}, and several classifications for the two-dimensional generalisation \cite{schuch-2006,verstraete-2006-96}. Developing QMA-complete instances of the translationally invariant local Hamiltonian problem will show that no approximation strategy (that is applicable to these cases) can give a certificate without first resolving the question QMA$\stackrel{?}{=}$P. This is in contrast to the previous discussions \cite{eisert-2006c,schuch-2006,verstraete-2006-96} which are tied to specific approximation strategies.

{\em Definitions:} For completeness, we repeat the following definitions:
\begin{definition}(QMA)
A promise problem $L$ with `yes' instances $L_{yes}$ and `no' instances $L_{no}$ is in QMA if there exists a quantum polynomial time verifier $V$ and polynomial $p$ such that for some $\epsilon(M)=2^{-\Omega(M)}$
\begin{itemize}
 \item $\forall x \in L_{yes}, \exists \ket{\xi} \quad \Pr \left( V(\ket{x},\ket{\xi} )=1\right)\geq 1-\epsilon$
 \item $\forall x \in L_{no}, \forall \ket{\xi} \quad \Pr \left( V(\ket{x},\ket{\xi})=1\right)\leq \epsilon$
\end{itemize}
where $|x|=M$, $\ket{\xi} \in {\cal B}^{\otimes p(M)}$ and $\Pr(V(\ket{x},\ket{\xi})=1)$ denotes the probability that $V$ outputs 1 given $\ket{x}$ and $\ket{\xi}$.
\end{definition}
\begin{definition}
Given a 2-local translationally invariant Hamiltonian on a one-dimensional chain of $N$ $D$-dimensional systems, $H=\sum_{j=1}^Nh_{j,j+1}$ with $\|h\|\leq\rm{poly}(N)$, and two constants $a<b$, the promise problem that we examine has yes instances in which the smallest eigenvalue of $H$ is at most $a$, and no instances in which it is larger than $b$, and we must decide which is the case. We call this problem TI2LH (Translationally Invariant 2-Local Hamiltonian).
\end{definition}

{\em Computation from a Hamiltonian:} The central part of previous proofs of QMA-completeness, as in this one, is the construction of Hamiltonians by showing how to implement an arbitrary quantum computation within the Hamiltonian (and hence can implement the verifier $V$). We shall assume that this computation acts on $N$ qubits, of which the first $M$ contain the string $x$ to be verified, and the other $N-M$ are initialised in $\ket{0}$, and are used as ancillas in the computation. This computation is typically achieved by introducing a `clock' system, and a Hamiltonian that implements a particular gate as the clock increments. The Hamiltonian hops the system through the different clock states, in a way that maps directly to quantum state transfer along a spin chain \cite{Bos03,Christandl,Kay:2004c}. In the original proposal \cite{KSV02a}, the hopping terms between neighboring clock states all had the same strengths. Since a uniformly coupled chain does not achieve perfect transfer of a state through the system \cite{Bos03,Kay:2004c} (and hence perfect arrival of the computation in the final state is impossible), the computation can be extended by a number of identity operations, thus ensuring that the probability that the computation has got past the final step is high \cite{AvK+a}. It has since been observed \cite{kay-2007} by making an analogy with perfect state transfer schemes \cite{Christandl,Kay:2004c} that varying the coupling strengths can allow perfect arrival of the computation, and hence the addition of further steps is unnecessary. However, here we will continue to use the uniformly coupled case due to the simplicity of the related eigenvectors \footnote{In particular, the lowest energy eigenvector of an $N$-qubit chain takes the form $\frac{1}{2^{N/2}}\sum_n(-1)^n\sqrt{\binom{N}{n}}\ket{n}$, and hence the final coefficient is exponentially suppressed}, particularly since for the proof of QMA-completeness, we do not actually perform the computation, and hence do not need perfect arrival of the result.

Ultimately, to implement a computation composed of a discrete set of gates, we need to introduce a clock system. However, in a translationally invariant Hamiltonian, we neither have time resolution nor spatial resolution in which to encode this clock. Instead, we choose to hold it in the state of each spin. Each of these $D$-dimensional spins will have a number of states: $\ket{on}$, and $\ket{x,y,z}$, where $x\in\{0,1\}$ is the qubit state which contains the computational component of the state, $z\in\{1,2\ldots N\}$ denotes the position of the spin in the lattice and $y\in\{0,1\ldots R\}$ is the time component, which we think of as `cycle number'. The input state to our computation will be a state of $N+1$ spins, $\ket{on}\bigotimes_{n=1}^{N}\ket{x_n,n,0}$ where $x_n$ are the bits of the string $x$ to be tested, and the $\ket{on}$ state indicates the initial position for a `read/write-head' that our translationally invariant Hamiltonian will be able to scan backwards and forwards across the chain in discrete cycles (it will always scan across all $N$ computational qubits). As it moves across, the cycle number is locally incremented by 1, and the required local unitaries can be implemented due to the position label. The computing part of the Hamiltonian takes the form
\begin{eqnarray}
&&h^{comp}_{i,j}=-\proj{on}_i\otimes\ket{0,1,0}\bra{0,1,1}_j							\nonumber\\
&&-U_{n,m}^{i,j}\otimes\ket{m,n}\bra{m-\delta_m,n}_i\otimes\ket{m,n'}\bra{m-1+\delta_m,n'}_j		\nonumber\\
&&-U_{N,m}^{i,j}\otimes\ket{m,N-1}\bra{m,N-1}_i\otimes\ket{m+1,N}\bra{m-1,N}_j			\nonumber\\
&&-U_{1,m}^{i,j}\otimes\ket{m+1,1}\bra{m-1,1}_i\otimes\ket{m,2}\bra{m,2}_j		+\text{h.c.}		\nonumber
\end{eqnarray}
where we are implicitly summing over cycle index $m$ and position index $n$ ($n'=n+1$, $2\delta_m=1+(-1)^m$). The unitaries $U_{n,m}^{i,j}$ implement the step of the computation that is required to be acted at time $m$ between qubits labelled $n$ and $n+1$, and only acts on the qubit space of sites $i$ and $j$. The first term initialises the read/write head. The next term propagates the head to the right or left depending on the value of $\delta$, and hence the parity of the cycle number. The final terms are responsible for reversing the direction of travel of the head when it reaches either extreme of the chain. For convenience, we fix the ground state energy of this Hamiltonian to zero by also adding the terms $$2\identity^{i}\otimes\proj{m,n}_i-\identity^i\otimes(\proj{0,1}+\proj{R,1})_i.$$
The final two terms mark a departure from the correspondence to transfer on a uniformly coupled chain, but makes the form of the lowest energy eigenvector of the Hamiltonian much simpler (see Eqn.~(\ref{eqn:evector})).

We now have a translationally invariant Hamiltonian $H_{comp}=\sum_ih_{i,i+1}$ where the terms $\|h\|\leq\rm{poly}(N)$. However, it currently acts on a non-translationally invariant input state. It is helpful to observe that the input states $\ket{\eta^n}$, where $\ket{on}$ is located on qubit $n$ can be combined to make the state translationally invariant, $\sum_n\ket{\eta^n}$. After computation, the result would still be available because projective measurement would locate the position of $\ket{on}$. Alternatively, one could use the elegant techniques described in \cite{vollbrecht-2005} to achieve computation on a translationally invariant input state.

{\em QMA-Completeness:} Now that we have a translationally invariant Hamiltonian for performing computations, we turn our attention to proving the QMA-completeness of TI2LH. We do this by adding some extra terms to the Hamiltonian
$$
H=J_1H_{input}+J_2(\alpha H_{form}+H_{comp})+R(N-1)H_{output}.
$$
Intuitively, the terms $H_{input}$ and $H_{form}$ will verify that we have the desired input by adding an energy penalty for incorrect states. $H_{input}$ is particularly used to ensure the state of ancillas for the computation. We take these to be positioned on qubits $M+1$ to $N$, and must be in the $\ket{0}$ state at time 0. Thus, we require an energy penalty
$$
h^{input}_i=\sum_{n=M+1}^N\proj{1,0,n}_i.
$$
The terms in $H_{form}$, while they could be applied just when the clock is at times 0, can equally well be applied at all time because they act on parts of the state that don't change, such as the $\ket{on}$ and position states. As such, they are more conveniently considered in conjunction with $H_{comp}$. The first term we apply is to ensure that there is at least one $\ket{on}$ present in the system, $-\proj{on}_i$. However, we don't want more than one of these states, so we also need an energy cost for an $\ket{on}$ being on the right of a state in position $n<N$, that more than compensates for the reduced energy due to the $\proj{on}$ term,
$$
\sum_{n<N}2\proj{x,m,n}_i\otimes\proj{on}_{i+1}.
$$
Similarly, we should ensure that neighboring position labels only increment by 1,
$$
\sum_{q\neq 1}\proj{x,m,n}_i\otimes\proj{x,m,n+q}_{i+1}.
$$
Equally, $H_{output}$ will test the final states, and add an energy penalty for a `no' result,
$$
h^{output}_i=\proj{1,R,1}_i,
$$
where the computation outputs the accept/reject code ($\ket{0}/\ket{1}$) on the first qubit after $R=\text{poly}(N)$ cycles. 

We must now verify that the eigenstates of $H$ satisfy the required properties of TI2LH. We choose to closely follow the proof presented in \cite{Kempe}, making use of their projection lemma,
\begin{lemma}
Let $H=H_1+H_2$ be the sum of two Hamiltonians operating on some Hilbert space $\cal H=\cal S+\cal S^\perp$. The
Hamiltonian $H_2$ is such that $\cal S$ is a zero eigenspace and the eigenvectors in $\cal S^\perp$ have eigenvalue at
least $J> 2\|H_1\|$. Then,
 $$\lambda(H_1|_{\cal S}) - \frac{\|H_1\|^2}{J-2\|H_1\|} \le \lambda(H) \le \lambda(H_1|_{\cal S}).$$
$\lambda(H)$ denotes the smallest eigenvalue of $H$.
\end{lemma}
In particular, one can select, say $J=8\|H_1\|^2+2\|H_1\|$ in order to provide the bounds
$$\lambda(H_1|_{\cal S}) - \frac{1}{8} \le \lambda(H) \le \lambda(H_1|_{\cal S}).$$

Firstly, let's consider that there are `yes' instances $x$ with bits $x_n$ (and $x_n=0$ if $n>M$). To test this solution, the input to the computation would have been
$$
\ket{\eta_0}=\ket{on}\bigotimes_{n=1}^N\ket{x_n,0,n}.
$$
We can consider the $(N-1)R$ discrete steps of the computation and write the state after each step as $\ket{\eta_n}$. Using these, we can write down a state
\begin{equation}
\ket{\eta}=\frac{1}{\sqrt{R(N-1)+1}}\sum_n\ket{\eta_n}	\label{eqn:evector}
\end{equation}
and evaluate that
$$
\bra{\eta}H_{input}\ket{\eta}=-1, \quad \bra{\eta}H_{comp}\ket{\eta}=0.
$$
Therefore, if the verifier accepts with probability $\geq 1-\epsilon$, $\bra{\eta}H\ket{\eta}\leq -J_1+\epsilon$.

Now we must prove that if there are no `yes' instances, all eigenvalues are larger than some value $b$. One can precisely follow the arguments of \cite{Kempe}. However, there is one small additional detail that we must first consider -- the energy gap between the ground-state space of $\alpha H_{form}+H_{comp}$ and the next excited state. Firstly we observe that due to $H_{form}$, the space splits into a series subspaces defined by the total number of $\ket{on}$ states present, and how badly ordered the position labels are. All of these subspaces have an offset of at least $\alpha$ from the original subspace of a single $\ket{on}$ state and correctly ordered position labels. Within this original subspace, we have the ground state subspace described by states $\ket{\eta}$ for all possible input values $x$, and cyclic permutations of the initial position of $\ket{on}$. Relative to the ground state space, all eigenvectors are separated by at least $c/(R(N-1))^2$ for some positive constant $c$ \cite{KSV02a}. This is readily proved by performing a transformation to the state transfer model. Thus, by setting $\alpha>c/(R(N-1))^2$, we can ensure that the minimum energy gap is at least $c/(R(N-1))^2$. We can now proceed with application of the projection lemma where $H_2=J_2(\alpha H_{form}+H_{comp})$ and $H_1=H-H_2$. As in \cite{Kempe}, we need to nest the application of this lemma, so that $H'=H_1|_{{\cal S}^{comp}}$, $H_2'=J_1H_{input}|_{{\cal S}^{comp}}$ and $H_1'=H'-H_2'$ before finally arriving at the result that $\lambda(H)\geq 3/4-\epsilon$, thereby proving that solving the local Hamiltonian problem implies the ability to solve all QMA problems.

We can transform this Hamiltonian of $N+1$ spins of dimension $D=2RN+1$ made up of two-body terms into a Hamiltonian acting on $N\log_2(D-1)+1$ copies of 7-level spins and $2\log_2(D-1)$-body terms. This transformation can be achieved by replacing each $D$-dimensional spin with a set of spins where one level is used as the $\ket{on}$ state, two levels encode a qubit state and the final three levels ($3\times 2+1=7$) denote whether that spin is being used to store a computational qubit, or a position or a cycle label, where an entire position and cycle label is achieved by combining $\log_2(D-1)$ neighbors. We require $N$ of these blocks, and a single extra spin to be initialised in the $\ket{on}$ state. The two-body terms from $H^{comp}$ need to act across two complete blocks of $\log_2(D-1)$ spins.

{\em Consequences and Conclusions:} In summary, if there exist `yes' instances to a QMA problem, the minimum eigenvalue of the corresponding $H$ (implementing the verifier $V$) is $\leq -J_1+\epsilon$, whereas if there are no such instances, the minimum eigenvalue is at least $-J_1+\smallfrac[3]{4}-\epsilon$. Therefore, solving TI2LH allows you to solve any QMA problem. Note that our construction of the Hamiltonian is necessarily degenerate \footnote{If we use open boundary conditions instead of periodic, the degeneracy due to permutation invariance is lifted.}, as seen by considering the case where there is a `yes' solution, tested by $\ket{\eta}$. There are $N+1$ equivalent states corresponding to the different possible starting positions of $\ket{on}$, all of the same energy. A similar argument will hold for the case where there are no such solutions. Our construction also implicitly shows how one can take a system on $N$ spins of dimension $d$ with $k$-body local Hamiltonian terms and make it into a Hamiltonian acting on $N$ spins of dimension $Nd$ with $k$-body local terms which is translationally invariant while preserving the structure of the eigenstates.

This proves that any attempt to classically approximate the ground state energy of a degenerate Hamiltonian of $N$ spins where either the Hamiltonian is composed of $O(\log(N))$-body (local) terms on spins of fixed dimension, or 2-body terms acting on spins of $\text{poly}(N)$ dimension is QMA-hard (in $N$) and thus, assuming QMA$\neq$P, will, in the worst case, be intractable. Specifically, by polynomially increasing the computational resources available for determining the ground state energy, the accuracy, in the worst case, cannot be more than polynomially increased. Falling within this class of Hamiltonians is the Bose-Hubbard model at constant filling fraction. At a filling fraction of 1, for example, if there are $N$ lattice sites, then there are $N$ bosons. We have to allow for the possibility that all $N$ bosons could be found in a single lattice site, and thus each lattice site must be treated as an $N$-level system.

As we have proven, while there is a degenerate ground state, there are also excited states of these Hamiltonians. Since there is an energy gap, properties such as exponentially decaying correlation functions \cite{hastings-2006-73} hold, and techniques such as DMRG/MPS, and Vidal's recently proposed Multi-scale Entanglement Renormalization Ansatz (MERA) \cite{vidal-2006a} provide an efficient representation. We conclude that finding this representation must be a computationally hard task. This does not say, of course, that classical approximation strategies can't be useful in a range of physically important problems since the QMA-hardness of a problem only elucidates the worst case scenario. In contrast to previous attempts at similar problems \cite{eisert-2006c,schuch-2006,verstraete-2006-96}, the example Hamiltonian that we have produced is not tied to a specific algorithm that one might be trying to implement, but instead applies to the general question that we are trying to solve with such an algorithm, thereby eliminating the possibility of globally efficient algorithms that encapsulate the Hamiltonians in question.

Evidently, an important question for the future is whether these results can be extended to the situation where both the local dimension and the Hamiltonian terms remain fixed as the system size grows, since this is the case which is most closely related to physical systems. While it seems unlikely that our technique can be directly applied (as we need to use a clock), there are strong indications that this may well be the case. In the first case we analysed, where the local Hilbert space dimension scales with $\text{poly}(N)$, but with two-body terms in the Hamiltonian, we know most of the structure of the ground state from Eqn.~(\ref{eqn:evector}) if there are `yes' solutions. By imposing this structure on the ansatz states of our approximation strategy, we are effectively reduced to minimising the energy over the local qubit states i.e.~the QMA-hardness of this problem is effectively encoded in a nearest-neighbor Hamiltonian acting on qubits.


We would like to thank Lluis Masanes and Jens Eisert for useful conversations. This work was supported by Clare College, Cambridge.

\end{document}